\documentclass[11pt]{article}
\usepackage{arxiv}

\usepackage{amsmath}  
\usepackage{graphicx}
\usepackage{filecontents}
\usepackage{multirow}
\usepackage{xcolor}
\usepackage{soul}
\usepackage{textcomp}
\title{Trees and Islands -- Machine learning approach to nuclear physics}
\author{
  Nishchal R. Dwivedi \\
 Nuclear Physics Division, Bhabha Atomic Research Centre, Trombay, Mumbai 400 085, India\\
 and\\
 Department of Physics, University of Mumbai, Vidyanagari Campus, Mumbai 400 098, India\\
 \\
  \texttt{dwivedi.nishchal@gmail.com} \\
  }

\begin{document}
\maketitle
\begin{abstract}
We implement machine learning algorithms to nuclear data. These algorithms are purely data driven and generate models that are capable to capture intricate trends. Gradient boosted trees algorithm is employed to generate a trained model from existing nuclear data, which is used for prediction for data of damping parameter, shell correction energies, quadrupole deformation, pairing gaps, level densities and giant dipole resonance for large number of nuclei. We, in particular, predict level density parameter for superheavy elements which is of great current interest. The predictions made by the machine learning algorithm is found to have standard deviation from 0.00035 to 0.73.

\end{abstract}


\section{Introduction}
Machine Learning (ML) and Artificial Intelligence (AI) have found a place in our contemporary technology to discover patterns, classify and predict any large data.
These classifications and predictions find applications from social media, online behaviour\cite{lee2010uncovering,perozzi2014deepwalk} to banking, stock market movement \cite{huang2005forecasting}, and so on. These algorithms are used by various social platforms to personalise our experience on the web, by suggesting us news\cite{liu2010personalized}, advertisements \cite{nath2013smartads} and pictures \cite{steels2006collaborative} of our interest. 
Recent use of ML in high energy physics for search of exotic particles \cite{baldi2014searching},  and making faster computations in molecular dynamics calculations \cite{botu2015adaptive} has opened the vistas of their usage in  fundamental sciences.

Due to the capability of harnessing of nuclear technology in medicine, agriculture, and clean energy, there is a need to have a collective database of various nuclear observables for research and diagnostic purposes. Organisations like IAEA have been maintaining the database as a standard and have a collection of observed and evaluated data of nuclei since late 1960's. 

ML prediction algorithms do not just provide a regression curve, but are capable of capturing more complex patterns in the data. In this letter we explore such an algorithm for model generation and prediction of nuclear observables. We use the available nuclear data and apply the Gradient Boosted Trees \cite{friedman2001greedy} prediction algorithm. This algorithm is trained on randomly sampled 60\% of the data, with a part of this 60\% used for validation. This training leads to a prediction model, which is generated as a regression tree. GBT minimizes the loss function of mean square error, and modifies the original model by correcting on the errors iteratively, till either the error minimises to a constant or the model starts to overfit. Overfitting is avoided by constantly checking the accuracy of the measurement from the validation set. The rest of the 40\% of data is used for testing the generated model. This ratio of 60\%-40\% is taken after optimisation and is found to work well with the considered observables. We define a standard deviation ($\sigma$), to quantify predictions, as the root mean square of difference between predicted values by the model and actual values given by the test data. Further, standard error is defined as the standard deviation divided by the square root of size of the sample. 

Recent works \cite{clark2006application,utama2016nuclear} use ML techniques to predict nuclear mass and charge radii using Support Vector Machines and Neural Networks. These works use two input parameters (\textit{features}) of number of neutrons and number of protons. These techniques though give a good predictions for the above observables, but they do not predict with similar efficiency for other nuclear data. Further, the ML algorithms learn better with more number of relevant \textit{features}. In our work, we do feature engineering for the problem and increase the number for \textit{features} from two to seven and then generate a prediction model for various nuclear data. We further employ the ML algorithm for two \textit{features} for the same set of data and compare the results in Table \ref{tab:comp}.

\section{Classification and prediction}
Consider a set of data where the input variables $x_{i,1},x_{i,2},\ldots,x_{i,j}$ (\textit{features}) give a resultant output $y_i$ through some process. The aim is to find a model $M$, such that, $M(x_{i,1},x_{i,2},\ldots,x_{i,j})=y_i$, where $i,j=1,2,\ldots$. In a typical ML scenario, a training set of a given data is used to teach the algorithm patterns in the output $y_i$'s. Based on this learning, the ML algorithm generates a model. This model can be a simple linear regression, a collection of trees (Random Forest), based on Neural networks, a classification of clustering (based on Support Vector machine), or any other of such techniques, depending on the problem. This generated model can then be used to predict $y_i$'s for $x_{i,j}$'s of interest, for which data is not even available.

Machine learning algorithms help in both classifying and predicting types of problems. In classification problems, the output $y_i$ is a discrete number, indicating classes or limited number of possible states where the data can exist. Here, a classification algorithm after learning gives the most probable class in which the state with a given set of features can exist in.

In predicting problems, the features correspond to a continuous value of $y_i$. Here, the algorithm has to learn through regression techniques to predict how the data trend will follow for the features of interest.

Gradient boosted trees (GBT) is a ML technique aimed at minimizing the loss function. Here, we demonstrate machine learning methods in nuclear physics using GBT.

\section{Calculations and results}
Modelling a nucleus and understanding its energy levels \cite{ajzenberg1976energy} has been of great interest.
The famous liquid drop model was proposed by Gamow and for which Bethe and von Weizs\"acker \cite{bethe1936nuclear,weizsacker1935theorie} proposed a semi empirical mass formula to describe a nucleus. This formula is written in terms of volume, surface, Coulomb interaction, mass asymmetry and pairing energies. In our approach, we treat these along with the number of protons and neutrons as the \textit{features} of the problem, as they are capable to model nuclear phenomena \cite{pomorski2004fission,reisdorf1994heavy}. After training the nuclear data, a model is generated which is used to give predictions on the test set, for which the actual values are known. If we plot the actual value vs the predicted values of the test set, the points should fall on the $y=x$ line, for perfect prediction.

\subsection{Shell correction energy and damping parameter, $\gamma$}
\begin{figure}[ht!]
\centering
\includegraphics{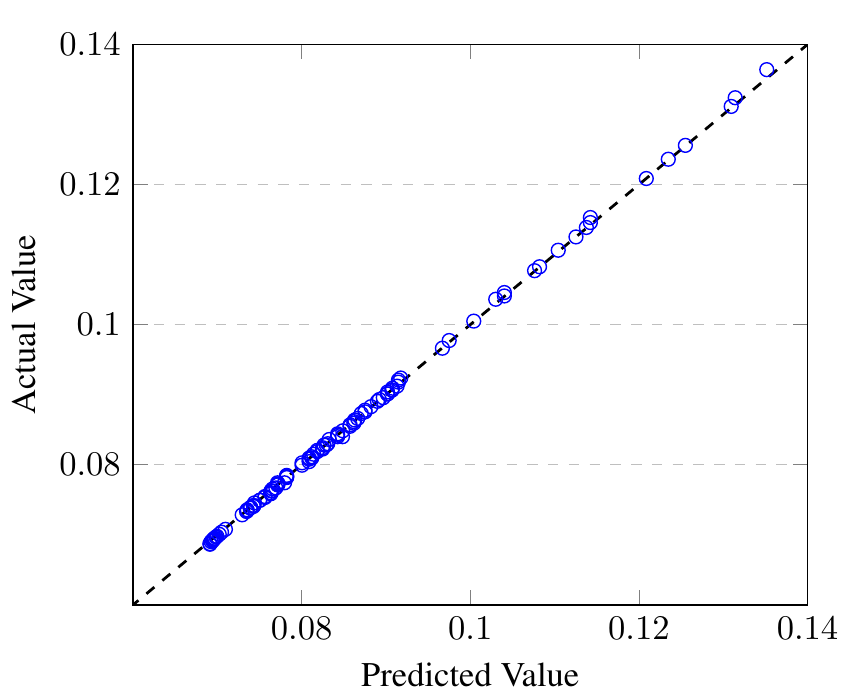}
\caption{Predictions for the test set for $\gamma$ values show a standard deviation of 0.00035 and a standard error of order $10^{-5}$. The data is trained and tested on a data set of 290 nuclei.}
\label{fig:gam}
\end{figure}

\begin{figure}[ht!]
\centering

\includegraphics{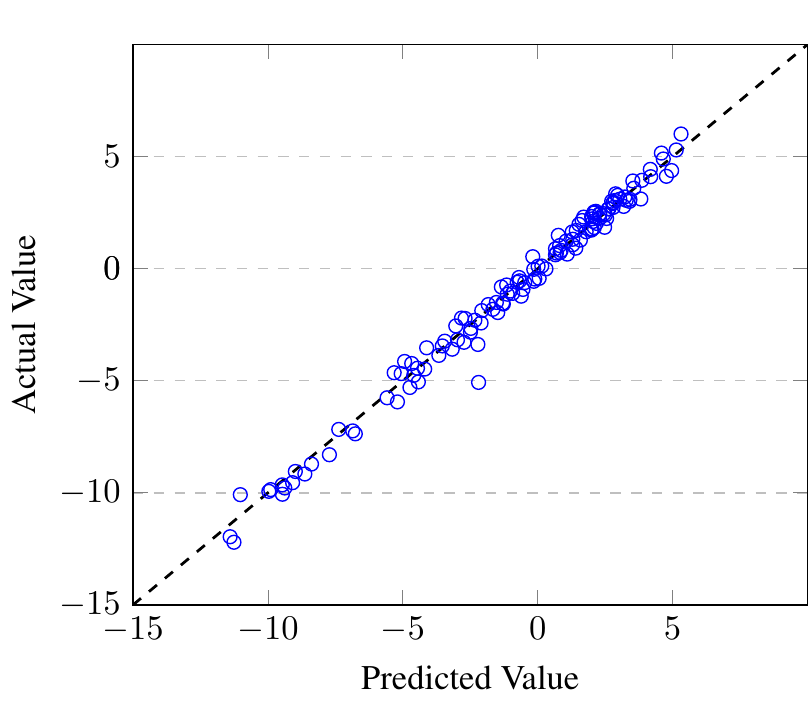}
\caption{The predictions of Shell correction energy give a standard deviation of 0.553 and standard error of 0.0067. The prediction model is made by training and testing on data set of 6735 nuclei.}
\label{fig:shl}
\end{figure}

Statistical models have been very useful to understand nuclei and their transitions. Usually, Fermi gas model \cite{bethe} is used to predict level density of the nucleus, but this model does not incorporate the shell effects. To incorporate shell corrections given by Strutinsky \cite{strutinsky1968shells}, phenomenological relations are used. One such popularly used relation is by Ignatyuk \cite{ignatyuk} as follows:

\begin{eqnarray}
a(E^{*})=\tilde{a}\left[1+\frac{\delta W}{E^{*}}\{1-\exp(-\gamma E^{*})\}\right]
\label{ign}
\end{eqnarray}

where $\tilde{a}$ is the asymptotic value of nuclear level density parameter. The damping coefficient, denoted by $\gamma$, is obtained by fitting gamma spectrum \cite{al2015thermal} and $\delta W$ denotes the shell correction energy which is given as the difference between the experimental binding energy of a nucleus and the binding energy calculated from the liquid drop model \cite{nndc}. The Gilbert-Cameron model \cite{gil} calculates $\gamma$ values by using neutron resonance data for various nuclei. This calculated data for 290 nuclei from $Z=$ 11 to 98 \cite{ripl} were used for $\gamma$ calculations. 6735 nuclei from $Z=8$ to 99 \cite{ripl} were used for $\delta W$ prediction. These values of $\delta W$ are calculated using  Mengoni-Nakajima mass formula \cite{mengoni1994fermi}. The prediction from this data for the damping coefficient is given by Fig. \ref{fig:gam} and for shell correction energy is given by Fig. \ref{fig:shl}. The prediction for $\gamma$ has the standard deviation as 0.00035, which shows an excellent prediction by the GBT method, and for $\delta W$ is 0.553, which shows a good prediction.

\subsection{$\beta_2$ deformation}
\begin{figure}[ht!]
\centering

\includegraphics{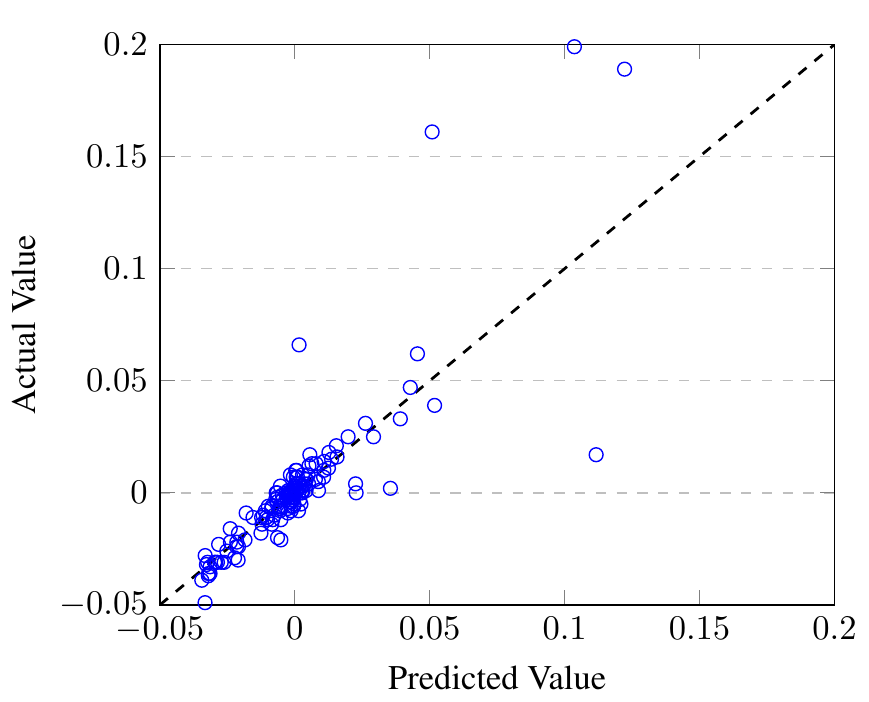}
\caption{Prediction on 8983 values of calculated Quadrupole deformations, $\beta_2$. It has the standard deviation of 0.015 and standard error of the order $10^{-4}$.}
\label{fig:b2}
\end{figure}

Understanding the shape and structure of a nucleus gives us information about the energy levels and transitions among them \cite{Bohr-Mottelson2}. Usually, the ground state deformation is calculated by using a finite -- range liquid drop model \cite{moller}. These have been tabulated in the RIPL library at \cite{ripl}. We use the quadrupole deformation values for 8983 nuclei from $Z=8$ to 136 from this library. We see that the test set values show a good prediction from the generated model with the standard deviation of 0.015 as seen in Fig. \ref{fig:b2}.

\subsection{Pairing gap}
\begin{figure}[ht!]
\centering

\includegraphics{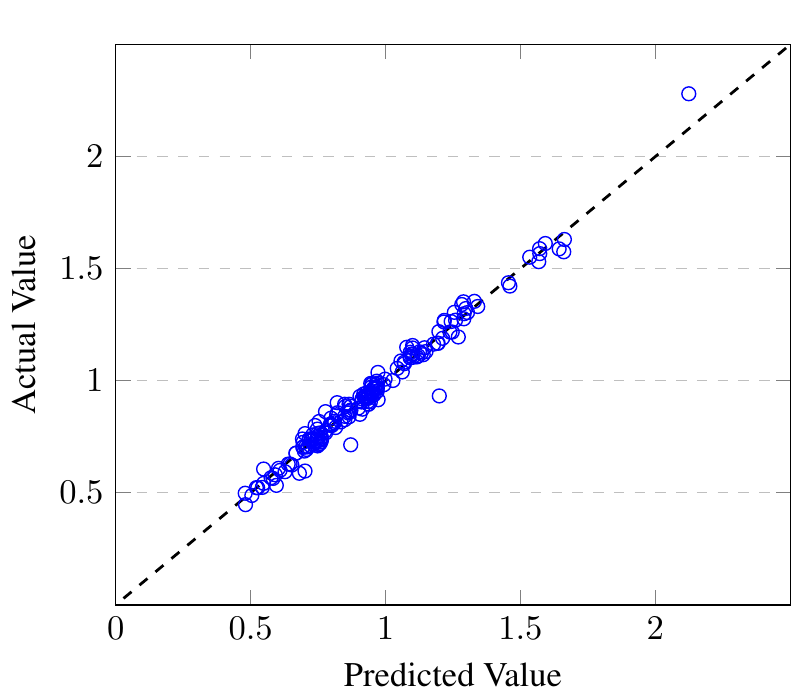}
\caption{Prediction of test set for pairing gaps for proton. The model is trained and tested on 8979 nuclei and give a standard deviation of 0.037 and standard error of the order $10^{-4}$.}
\label{fig:protonpg}
\end{figure}

\begin{figure}[ht!]
\centering

\includegraphics{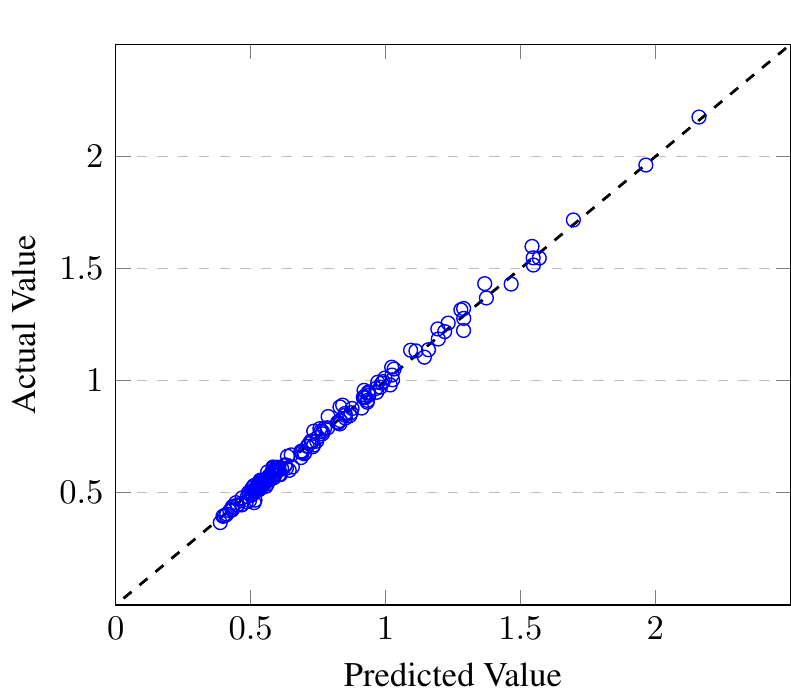}
\caption{Prediction of test set for pairing gaps for neutron. The model is trained and tested on 8979 nuclei and give a standard deviation of 0.037 and standard error of the order $10^{-4}$.}
\label{fig:neutronpg}
\end{figure}

The possibility of existence of pairing in the nuclei was postulated by Bohr and others \cite{bohrpines}, which was a success in explaining the dependence of the binding energy of the nucleus with the even and odd number of protons and neutrons. The pairing gap calculated using BCS pairing model with energy levels obtained by the folded-Yukawa single-particle model \cite{mollerpair} for 8979 nuclei from $Z=$ 8 to 136 \cite{lanl} is used. The predictions for the test set for proton and neutron pairing gaps are shown in Fig. \ref{fig:protonpg} and Fig. \ref{fig:neutronpg} respectively with their respective standard deviations as 0.037 and 0.023.

\subsection{Level density parameter}

Level density parameter (LDP) is the most important extracted quantity to understand nuclear observables like neutron resonances and reaction cross sections \cite{lang}. Recent semiclassical trace formula (STF) \cite{kaur} evaluates the temperature- dependent LDP within 10\%-15\% of the experimental values for magic and semi magic nuclei. This trace formula models nucleus using Harmonic oscillator \cite{brack1995analytical} with spin-orbit interactions \cite{amann2002semiclassical}. This formalism also incorporates the shell effects in the nuclei \cite{brack1972funny}. 

The Gilbert-Cameron (GC) model fits the Fermi gas model with the neutron resonances to obtain asymptotic LDP. These LDP are tabulated in \cite{ripl} for 289 nuclei from $Z=11$ to 98.

\begin{figure}[ht!]
\centering

\includegraphics{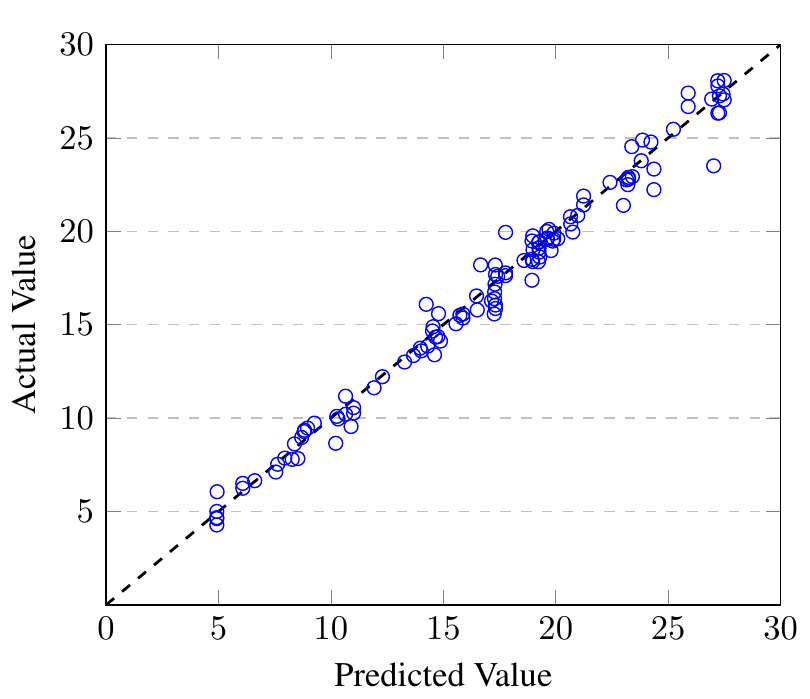}
\caption{By training and testing Level Density Parameter by the GC model for 289 nuclei, we get the standard deviation of 0.73 and standard error of 0.042.}
\label{fig:LDPGC}
\end{figure}

The results from GC show (Fig. \ref{fig:LDPGC}) a standard variation of 0.73 and a good agreement of the asymptotic level densities.

\begin{figure}[ht!]
\centering
\includegraphics{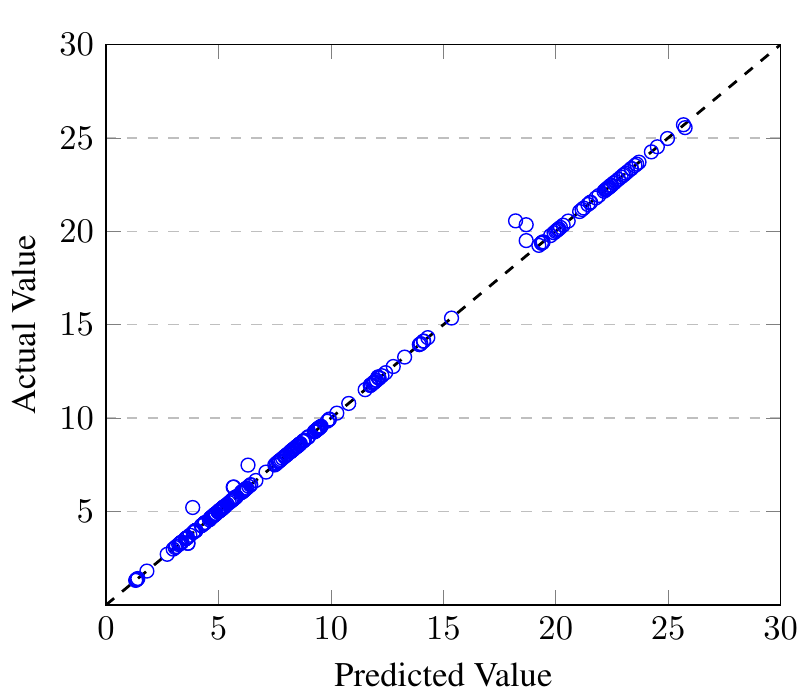}
\caption{Training and testing on the temperature dependent Level Density Parameter by Semiclassical trace formula for 32 nuclei at various temperatures (3000 data points) gives the standard deviation of 0.3 and standard error of 0.005. Comparison with Fig. \ref{fig:LDPGC} shows that the model trained with STF gives better predictions.}
\label{fig:LDPSTF}
\end{figure}

The STF is an exact formula with no adjustable parameters and this formalism also gives level density as a function of temperature. We train the algorithm for temperature dependent level density for 32 nuclei for various temperatures. We have 3000 data points. Temperature is added as a feature in this case along with the liquid drop energies and number of neutrons and protons. The ML generated model shows a standard deviation of 0.3 and the predicted values show a good agreement with the actual values (Fig. \ref{fig:LDPSTF}).

We further venture towards the \textit{island} of stability and predict the LDP of superheavy elements.

\subsection{Superheavy elements}

The synthesis of superheavy elements (SHE) $(Z>104)$ is a big challenge in nuclear physics and is of great contemporary interest due to its importance in heavy ion fusion reactions \cite{hofmann2000discovery}. The variation of LDP with the number of nucleons is generally seen to have an $A/n$ trend, where $A$ is the mass number of the nucleus and $n$ is the positive real number. Calculations suggest that for SHE, $n$ is about 11-13 \cite{bezbakh2015level}. We use the ML models generated from GC and STF to predict the level densities of some super heavy elements as in Table \ref{tab:SHE} and calculate the respective $n$- values as $n_{GC}$ and $n_{STF}$. We find that the STF does better than GC as expected, due to its exact nature, more information on temperature dependence and incorporation of spin orbit interaction in the trace formula.

\begin{table}[ht!]
    \centering
    \begin{tabular}{|c|c|c|c|c|c|}
 \hline
 Element & Z & GC & $n_{GC}$ & STF &$n_{STF}$ \\ 
 \hline
 $^{267}Rf$ & 104 &27.44 & 9.73   & 19.85     & 13.45\\
 \hline
 $^{268}Db$ & 105& 27.62 & 9.70   & 19.88 & 13.48\\ 
 \hline
 $^{269}Sg$ & 106 & 27.67 & 9.72   &  19.98    &13.46\\
 \hline
 $^{270}Bh$ & 107& 24.51 &9.81   & 19.98 &13.51\\ 
 \hline
 $^{270}Hs$ & 108& 27.10 &11.20   & 19.98       &13.51\\ 
 \hline
 $^{278}Mt$ & 109& 27.40 &10.14   &  19.97    &13.92\\ 
 \hline
 $^{281}Ds$ & 110& 27.44 &10.21   &  19.97    &14.07\\ 
 \hline
 $^{282}Rg$ & 111& 27.40 &10.29   & 19.97    &14.12\\ 
 \hline
 $^{285}Cn$ &112 & 27.44 &10.39   & 19.94    &14.29\\ 
 \hline
 $^{286}Nh$ & 113& 27.40 &10.43   &  19.95    &14.33\\ 
 \hline
 $^{289}Fl$ & 114& 27.44 &10.53   &  19.93    &14.50\\ 
 \hline
 $^{290}Mc$ & 115& 27.62 &10.50   &  19.93   &14.55\\ 
 \hline
 $^{293}Lv$ & 116& 27.44 &10.67   &  19.92  &14.70\\ 
 \hline
 $^{294}Ts$ & 117& 27.66 &10.62   &  19.94   &14.75\\ 
 \hline
 $^{294}Og$ & 118& 27.55 &10.67   &  19.94        &14.75\\ 
 \hline
 $^{315}Uue$ & 119& 27.44 &11.48   &  19.93  &15.81\\ 
 \hline
 $^{299}Ubn$ & 120& 27.55 &10.85   & 20.22    &14.79\\ 
 \hline
 $^{320}Ubu$ & 121& 27.40 &11.68   &  19.96    &16.04\\ 
 \hline
 $^{306}Ubb$ & 122& 27.43 &11.15   & 21.26     &14.39\\ 
 \hline
 
\end{tabular}
    \caption{Level desities predictied for the Super Heavy elements by GC model based on fermi gas model and semiclassical trace formula (STF). The values are close to the trend of $A/10$ for GC and $A/(13-14)$ for STF.}
    \label{tab:SHE}
\end{table}

\subsection{Giant Dipole Resonance}

During a photonuclear reaction, signatures of the nuclei behaving like a giant dipole with collective oscillations \cite{clark2001isoscalar} are observed. These oscillations are called as Giant Dipole Resonance (GDR). It corresponds to the fundamental absorption frequency of electric dipole radiation of the nucleus acting as a whole. It is often simply understood as oscillations in the nucleus of the neutrons against the protons. GDR dominates the photo-absorption process at 10-30 MeV of energies \cite{berman1975measurements}.

 We take the experimentally observed values for 180 nuclei for their observed first peak of energy corresponding to GDR \cite{ripl}. We then train our model and predict for a randomly selected test set. These experimental values give predictions with standard deviation of 0.76. The experimental standard deviation of these values from various experiments lie between 0.01 to 11.74.

\begin{figure}[ht!]
\centering

\includegraphics{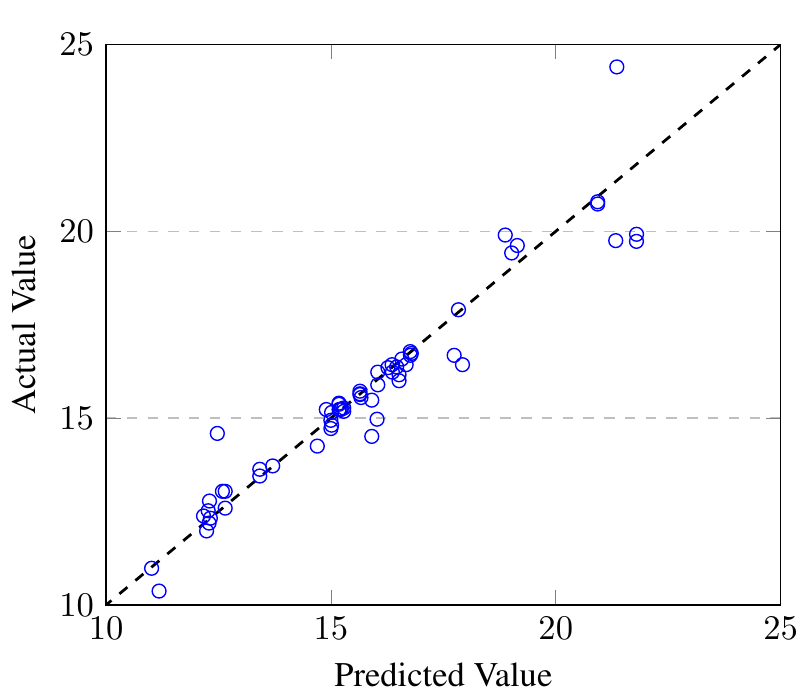}
\caption{Training and testing the experimental first peak energy values for GDR for 180 nuclei. It shows the standard deviation of 0.76 and standard error of 0.056.}
\label{fig:GDR}
\end{figure}

\begin{table}[ht!]
    \centering
    \begin{tabular}{c|c|c}
 \hline
 Observable & $\sigma$ (2 features) & $\sigma$ (7 features) \\ 
 \hline
 \hline
 Damping parameter & 0.00170 & 0.00035 \\
 \hline
 Shell Correction Energy & 0.910& 0.553\\ 
 \hline
 $\beta_2$ deformation & 0.018 & 0.015\\
 \hline
 Proton pairing gap & 0.040& 0.037\\ 
 \hline
 Neutron pairing gap & 0.640& 0.037\\ 
 \hline
 LDP (GC) & 0.790& 0.730\\ 
 \hline
 LDP (STF) & 0.600& 0.300\\ 
 ~&(3 features) &(8 features)\\
 \hline
 
\end{tabular}
    \caption{The table shows a comparison of standard deviations ($\sigma$) obtained after predictions on the test set on various nuclear data by using (1) two features (number of neutrons and protons) and (2) seven features (volume, surface, Coulomb interaction, mass asymmetry, pairing energies along with number of neutrons and protons). The value of $\sigma$ for ideal predictions is zero. The predictions made by using seven features show lower $\sigma$ values and hence they capture the trends in nuclear data in a better manner than two features. The LDP (STF) being temperature dependent level density has an extra feature of temperature.} 
    \label{tab:comp}
\end{table}

\section{Conclusion}
Nuclear data has been studied extensively using a machine learning algorithm. Phenomenological quantities like damping factor and shell correction energies, quantities with quantum mechanical origin like pairing gaps, evaluated values of quadrupole deformations, statistical quantities of level densities and experimental quantities of first peak energy of giant dipole resonances have been modelled. The predictions from these models show a very good agreement with the actual values with standard deviations ranging from 0.00035 to 0.73. To compare with contemporary literature, similar investigations on time series analysis \cite{singh2007application}, molecular physics \cite{rupp2012fast}, astrophysics \cite{solomatine2009novel}, economics \cite{park2015using,gabralla2013oil} and climate studies \cite{deo2015application} report  standard deviations ranging between 0.02 to 75.0. Our results, thus, are well acceptable, particularly as they are better by two orders of magnitude for damping parameter, $\gamma$.
We use these predictions to calculate the level density parameter of the superheavy elements in the island of stability, the values obtained sharpen the expected values from proposed theories.  We also show how the engineered \textit{features} increase the accuracy of the predictions by the generated models.

These algorithms are based and derived only from the data and not from nuclear theoretical modelling or phenomenological analysis of the experiments. Though the theoretical basis of these algorithms are statistical in nature, they capture the patterns in the data very well. This is somewhat reminiscent of the success of ideas originating in random matrix theory which work very well in nuclear physics, quantum chaos and so on \cite{mehta2004random}. We hope that this work will pave the way for further investigations using machine learning and AI as tools in the fields of physics where a lot of data is available while a clear cut theory is not yet available.   

Alternatively, the limited number of \textit{features} in the problem and good predictions, makes ML a good tool to get an insight for global mathematical modelling of nuclei, which is valid for all ranges of mass numbers. ML can be instrumental in understanding nuclear properties for which the experiments are not yet feasible.


\section{Acknowledgements}
The author expresses his deep gratitude to Prof. Sudhir Ranjan Jain (BARC) for discussions which led to this problem. The author is grateful to Aayushi Gupta (Indira Gandhi Institute of Development Research, Mumbai), Raman Sehgal (BARC) and Dr. Kaushal Sharma (Inter-University Centre for Astronomy and Astrophysics, Pune) for involved discussions.

The author acknowledges gratefully the funding for research from the Department of Atomic Energy and University of Mumbai - Bhabha Atomic Research Centre collaboration.


\end{document}